\documentclass{cernphprep}

\usepackage{times}
\usepackage{graphicx}
\usepackage{bm}
\usepackage{cite}

\usepackage{color}

\begin{document}

\begin{titlepage}

\EXPnumber{DIRAC/PS212}
\PHnumber{2016-128}
\PHdate{\today}

\title{Observation of $\pi^- K^+$ and $\pi^+ K^-$ atoms}

\begin{Authlist}

B.~Adeva\Iref{s}, 
L.~Afanasyev\Iref{d}, 
Y.~Allkofer\Iref{zu}, 
C.~Amsler\Iref{be}, 
A.~Anania\Iref{im},
S.~Aogaki\Iref{b},
A.~Benelli\Iref{cz}, 
V.~Brekhovskikh\Iref{p},  
T.~Cechak\Iref{cz}, 
M.~Chiba\Iref{jt}, 
P.~Chliapnikov\Iref{p},   
P.~Doskarova\Iref{cz}, 
D.~Drijard\Iref{c},
A.~Dudarev\Iref{d}, 
D.~Dumitriu\Iref{b}, 
D.~Fluerasu\Iref{b}, 
A.~Gorin\Iref{p}, 
O.~Gorchakov\Iref{d},
K.~Gritsay\Iref{d}, 
C.~Guaraldo\Iref{if}, 
M.~Gugiu\Iref{b}, 
M.~Hansroul\Iref{c}, 
Z.~Hons\Iref{czr}, 
S.~Horikawa\Iref{zu},
Y.~Iwashita\Iref{jk},
V.~Karpukhin\Iref{d}, 
J.~Kluson\Iref{cz}, 
M.~Kobayashi\Iref{k}, 
V.~Kruglov\Iref{d}, 
L.~Kruglova\Iref{d}, 
A.~Kulikov\Iref{d}, 
E.~Kulish\Iref{d},
A.~Kuptsov\Iref{d}, 
A.~Lamberto\Iref{im}, 
A.~Lanaro\Iref{u},
R.~Lednicky\Iref{cza}, 
C.~Mari\~nas\Iref{s},
J.~Martincik\Iref{cz},
L.~Nemenov\IIref{d}{c},
M.~Nikitin\Iref{d}, 
K.~Okada\Iref{jks}, 
V.~Olchevskii\Iref{d}, 
M.~Pentia\Iref{b}, 
A.~Penzo\Iref{it}, 
M.~Plo\Iref{s},
P.~Prusa\Iref{cz},  
G.~Rappazzo\Iref{im}, 
A.~Romero Vidal\Iref{if},
A.~Ryazantsev\Iref{p},
V.~Rykalin\Iref{p},
J.~Saborido\Iref{s}, 
J.~Schacher\IAref{be}{*},
A.~Sidorov\Iref{p}, 
J.~Smolik\Iref{cz}, 
F.~Takeutchi\Iref{jks}, 
L.~Tauscher\Iref{ba},
T.~Trojek\Iref{cz}, 
S.~Trusov\Iref{m}, 
T.~Urban\Iref{cz},
T.~Vrba\Iref{cz},
V.~Yazkov\Iref{m}, 
Y.~Yoshimura\Iref{k}, 
M.~Zhabitsky\Iref{d}, 
P.~Zrelov\Iref{d} 

\end{Authlist}

\Instfoot{s}{Santiago de Compostela University, Spain}
\Instfoot{d}{JINR, Dubna, Russia}
\Instfoot{zu}{Zurich University, Switzerland}
\Instfoot{be}{
Albert Einstein Center for Fundamental Physics, 
Laboratory of High Energy Physics, Bern, Switzerland
}
\Instfoot{im}{INFN, Sezione di Trieste and Messina University, Messina, Italy}
\Instfoot{b}{
IFIN-HH, National Institute for Physics and Nuclear Engineering, 
Bucharest, Romania
}
\Instfoot{cz}{Czech Technical University in Prague, Czech Republic}
\Instfoot{p}{IHEP, Protvino, Russia}
\Instfoot{jt}{Tokyo Metropolitan University, Japan}
\Instfoot{c}{CERN, Geneva, Switzerland}
\Instfoot{if}{INFN, Laboratori Nazionali di Frascati, Frascati, Italy}
\Instfoot{czr}{Nuclear Physics Institute ASCR, Rez, Czech Republic}
\Instfoot{jk}{Kyoto University, Kyoto, Japan}
\Instfoot{k}{KEK, Tsukuba, Japan}
\Instfoot{u}{University of Wisconsin, Madison, USA} 
\Instfoot{cza}{Institute of Physics ASCR, Prague, Czech Republic}
\Instfoot{jks}{Kyoto Sangyo University, Kyoto, Japan}
\Instfoot{it}{INFN, Sezione di Trieste, Trieste, Italy}
\Instfoot{ba}{Basel University, Switzerland}
\Instfoot{m}{
Skobeltsin Institute for Nuclear Physics of Moscow State University, 
Moscow, Russia
}

\Anotfoot{*}{Corresponding author} 

\Collaboration{DIRAC Collaboration}
\ShortAuthor{DIRAC Collaboration}

\newpage

\begin{abstract}
The observation of hydrogen-like $\pi K$ atoms, 
consisting of $\pi^- K^+$ or $\pi^+ K^-$ mesons, 
is presented. The atoms have been produced by 
24 GeV/$c$ protons from the CERN PS accelerator,  
interacting with platinum or nickel foil targets. 
The breakup (ionisation) of $\pi K$ atoms in 
the same targets yields characteristic $\pi K$ pairs, 
called ``atomic pairs'', with small relative momenta 
in the pair centre-of-mass system. 
The upgraded DIRAC experiment has observed $349\pm62$ 
such atomic $\pi K$ pairs, corresponding to a signal of 
5.6 standard deviations. 
\end{abstract}
\vspace{2cm}
\end{titlepage}

\section{Introduction}
\label{sec:intro}

Up to now, the DIRAC collaboration has published indications about 
the production of $\pi K$ atoms\footnote{The term $\pi K$ atom or  
$A_{K \pi}$ refer to $\pi^- K^+$ and $\pi^+ K^-$ atoms.} 
\cite{ADEV09,ALLK08,ADEV14}. 
This time, DIRAC reports the first statistically significant 
observation of the strange dimesonic $\pi K$ atom. 

Meson-meson interactions at low energy are the simplest 
hadron-hadron processes and allow to test low-energy QCD, 
specifically Chiral Perturbation Theory (ChPT) 
\cite{WEIN66,GASS85,MOUS00,COLA01}. The observation and 
lifetime measurement of $\pi^+\pi^-$~atoms (pionium) 
have been reported in \cite{AFAN93,ADEV05,ADEV11}. 
Going one step further, the observation and lifetime measurement 
of the $\pi K$ atom involving strangeness 
provides a direct determination of 
a basic S-wave $\pi K$ scattering length difference \cite{BILE69}. 
This atom is an electromagnetically bound $\pi K$ state 
with a Bohr radius of $a_{B}$~=~249~fm and 
a ground state binding energy of $E_{B}$~=~2.9~keV. 
It decays predominantly by strong interaction into 
two neutral mesons $\pi^0 K^0$ or $\pi^0 \bar{K^0}$. 
The atom decay width~$\Gamma_{\pi K}$ in the ground state (1S)   
is given by the relation \cite{BILE69,SCHW04}: 
$\Gamma_{\pi K} = \frac{1}{\tau} = R (a_{0}^-)^2$, 
where $a_{0}^-=\frac{1}{3}(a_{1/2}-a_{3/2})$ is 
the S-wave isospin-odd $\pi K$ scattering length 
($a_I$ is the $\pi K$ scattering length for isospin $I$) and 
$R$ a precisely known factor (relative precision 2\%). 
The scattering length $a_{0}^-$ has been studied in ChPT 
\cite{BERN91,KUBI02,BIJN04}, in the dispersive framework  
\cite{BUET04} and in lattice QCD (see e.g. \cite{SASA14}). 
Using $a_{0}^-$ from \cite{BUET04}, one predicts for the 
$\pi K$ atom lifetime $\tau = (3.5 \pm 0.4)\cdot10^{-15}~\text{s}$. 

\begin{figure}[ht]
\begin{center}
\includegraphics[width=0.4\columnwidth]{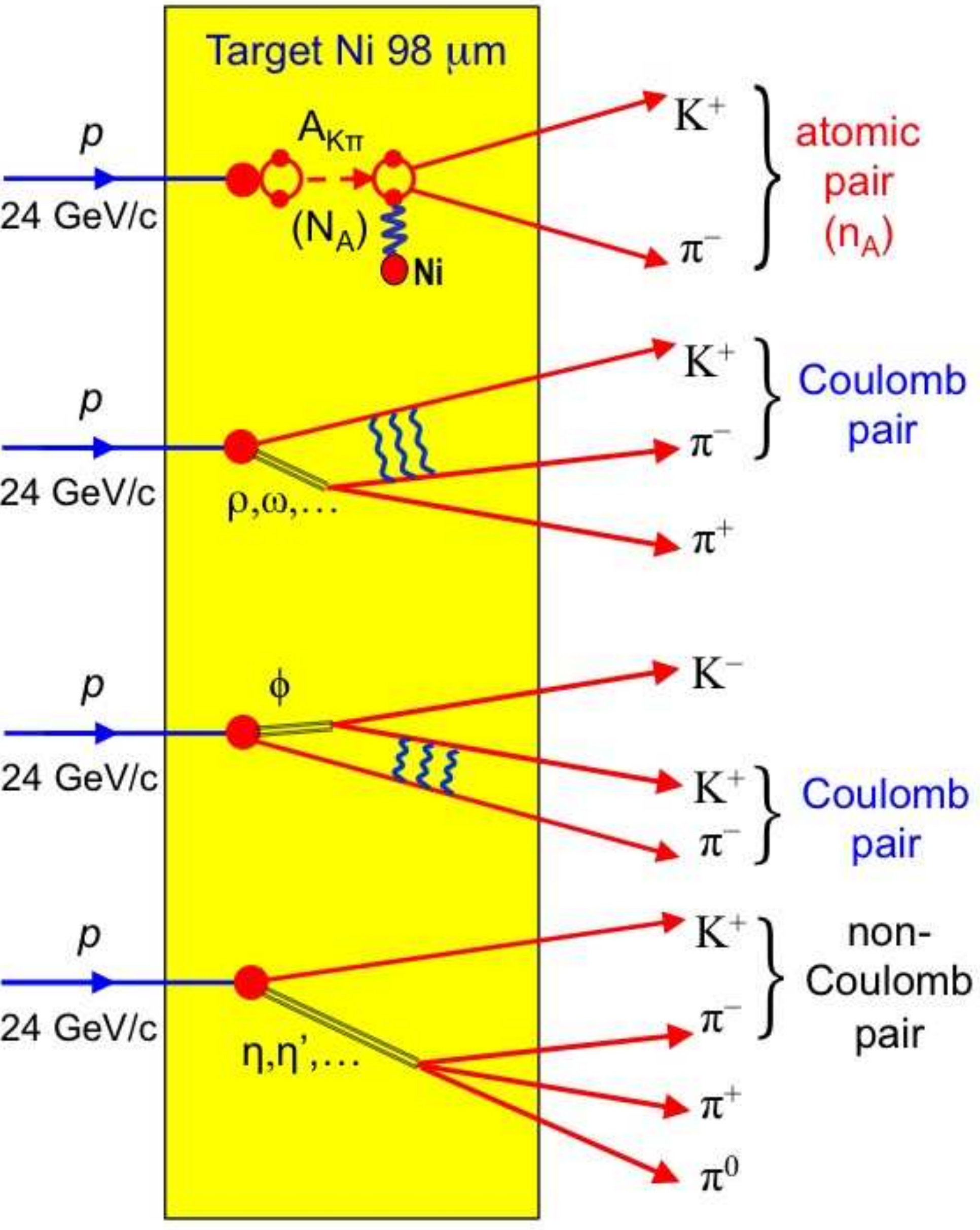}
\caption{Inclusive $\pi K$ production in  
24~GeV/c p-Ni interaction: 
p + Ni $\to$ $\pi^- K^+$ + X. 
The ionisation or breakup of $A_{K \pi}$ leads to 
so-called atomic pairs. 
(More details, see text in section~\ref{boundfree}.)}
\label{Fig1_1}
\end{center}
\end{figure}

A method to produce and observe hadronic atoms has been 
developed \cite{NEME85}. In the DIRAC experiment, 
relativistic dimesonic bound states, 
formed by Coulomb final state interaction (FSI), 
are moving inside the target and can break up. 
Particle pairs from breakup (atomic pair in Fig.~\ref{Fig1_1}) 
are characterised by a small relative momentum $Q <$ 3~MeV/c 
in the centre-of-mass (c.m.) system of the pair\footnote{
The quantity $Q$ denotes the experimental c.m. relative momentum. 
The longitudinal ($Q_L$) and transverse $(Q_T~=~\sqrt{Q^2_X + Q^2_Y})$ 
components of the vector $\vec Q$ are defined with respect to 
the direction of the total laboratory pair momentum.}. 

A first $\pi K$ atom investigation has been  
performed with a platinum target at the CERN PS with 
24~GeV/$c$ protons in 2007 \cite{ADEV09,ALLK08}. 
An enhancement of $\pi K$ pairs at low relative momentum 
has been observed, corresponding to $173 \pm 54$ 
$\pi K$ atomic pairs or a significance of 
3.2 standard  deviations ($\sigma$). 
In the experiment from 2008 to 2010, 
DIRAC has detected in a Ni target an excess of $178\pm49$ 
$\pi K$ pairs, an effect of only 3.6 $\sigma$ \cite{ADEV14}. 

In the present paper, experimental data obtained in Ni and 
Pt targets have been analysed, using recorded informations 
from all detectors (see Fig.~\ref{Fig2_1}) and enhanced 
background description based on Monte Carlo (MC) simulations.  
Setup geometry correction, detector response simulation, 
background suppression and admixture evaluation have been 
significantly improved for all runs. 

The above mentioned improvements allow a statistically 
reliable observation of $\pi K$ atoms.

\section{Experimental setup}

\begin{figure}[ht]
\begin{center}
\includegraphics[width=0.9\columnwidth]{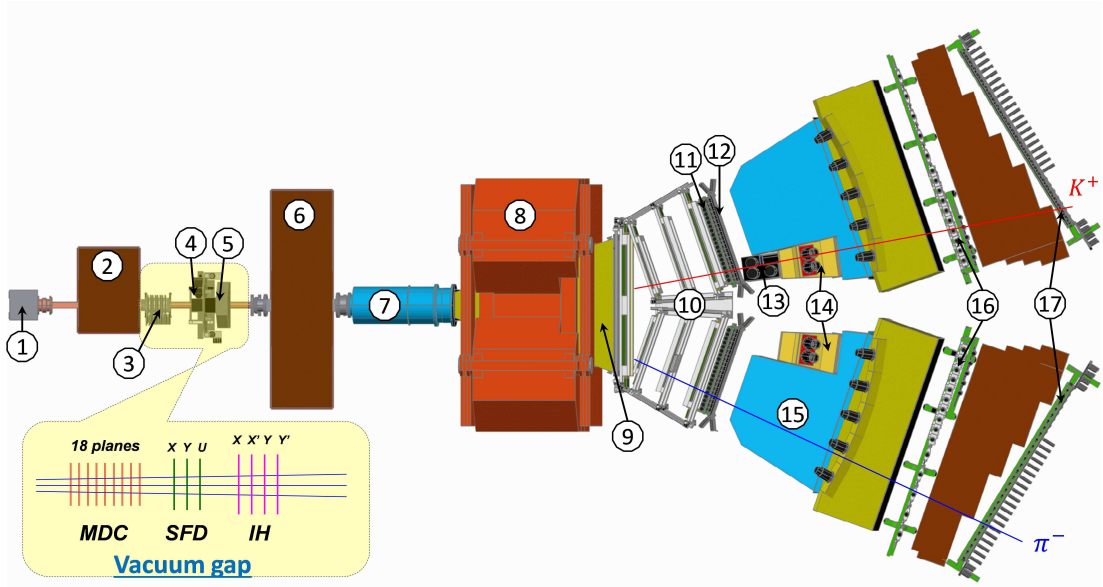}
\caption{ General view of the DIRAC setup:  
1 -- target station;
2 -- first shielding wall;
3 -- microdrift chambers;
4 -- scintillating fiber detector; 
5 -- ionisation hodoscope; 
6 -- second shielding wall; 
7 -- vacuum tube; 
8 -- spectrometer magnet; 
9 -- vacuum chamber; 
10 -- drift chambers; 
11 -- vertical hodoscope; 
12 -- horizontal hodoscope; 
13 -- aerogel Cherenkov; 
14 -- heavy gas Cherenkov; 
15 -- nitrogen Cherenkov; 
16 -- preshower; 
17 -- muon detector.}
\label{Fig2_1}
\end{center}
\end{figure}

The setup \cite{DIRA14}, sketched in Fig.~\ref{Fig2_1}, 
detects and identifies $\pi^+ \pi^-$, $\pi^- K^+$ and 
$\pi^+ K^-$ pairs with small $Q$. The structure of 
these pairs after the magnet is approximately 
symmetric for $\pi^+ \pi^-$ and asymmetric 
for $\pi K$. Originating from a bound system, 
these particles travel with the nearly same velocity, and 
hence for $\pi K$ atomic pairs, the kaon momentum is 
by a factor of about $\frac{M_{K}}{M_{\pi}} \approx 3.5$ larger 
than the pion momentum ($M_{K}$ is the charged kaon mass and 
$M_{\pi}$ the charged $\pi$ mass). 
The 2-arm vacuum magnetic spectrometer presented 
is optimized for simultaneous detection of 
these pairs \cite{note0505,note0523}.

The 24~GeV/c primary proton beam, extracted from the CERN PS, 
hits a ($26\pm1$)~$\mu$m thick Pt target in 2007\footnote{
The Pt target maximizes production of atomic pairs.} 
and Ni targets with thicknesses ($98\pm1$)~$\mu$m   
in 2008 and ($108\pm1$)~$\mu$m in 2009 and 2010\footnote{
The Ni targets are optimal for lifetime measurement.}. 
The radiation thickness of the 98 (108)~$\mu$m Ni 
target amounts to about $7\cdot 10^{-3}$ 
 $X_{0}$ (radiation length). 

The secondary channel (solid angle $\Omega = 1.2 \cdot 10^{-3}$~sr) 
together with the whole setup is vertically inclined 
relative to the proton beam by $5.7^\circ$ upward.  Secondary 
particles are confined by the rectangular beam collimator inside of 
the second steel shielding wall, and the angular divergence in  
the horizontal (X) and vertical (Y) planes is $\pm 1^\circ$.
 With a spill duration of 450~ms, the beam intensity has been  
(10.5--12)$\cdot 10^{10}$ protons/spill and, correspondingly,   
the single counting rate in one plane of the ionisation 
hodoscope (IH) (5--6)$\cdot 10^6$ particles/spill. 
Secondary particles propagate mainly in vacuum up 
to the Al foil at the exit of the vacuum chamber, 
which is located between the poles of the dipole magnet 
($B_{max}$ = 1.65~T and $BL$ = 2.2~T$\cdot$m). In the vacuum gap, 
MicroDrift Chambers (MDC) with 18 planes and  
a Scintillating Fiber Detector (SFD) with 3 planes 
X, Y and U, inclined by $45^\circ$, 
have been installed to measure particle coordinates   
($\sigma_{SFDx} = \sigma_{SFDy} = 60~\mu$m, 
$\sigma_{SFDu} = 120~\mu$m) and particle time   
($\sigma_{tSFDx} = 380$~ps, $\sigma_{tSFDy} = \sigma_{tSFDu} = 520$~ps).
The four IH planes serve to identify unresolved double track events 
with only one hit in SFD.

Each spectrometer arm is equipped with the following subdetectors: 
drift chambers (DC) to measure particle coordinates with 
about 85~$\mu$m precision; vertical hodoscope (VH) 
to measure time with 110~ps accuracy for particle identification 
via time-of-flight determination; horizontal hodoscope (HH) 
to select pairs with vertical separation less than 75 mm 
between the arms ($Q_{Y}$ less than 15 MeV/c);  
aerogel Cherenkov counter (ChA) to distinguish kaons from protons; 
heavy gas ($C_{4}F_{10}$) Cherenkov counter (ChF) to distinguish 
pions from kaons; nitrogen Cherenkov (ChN) and preshower 
(PSh) detector to identify $\mathrm{e}^+\mathrm{e}^-$~pairs; 
iron absorber and two-layer scintillation counter (Mu) 
to identify muons. In the ``negative'' arm, an aerogel counter 
has not been installed, because the number of antiprotons is small 
compared to $K^{-}$.

Pairs of oppositely charged time-correlated particles (prompt pairs) 
and accidentals in the time interval $\pm 20$~ns are selected by 
requiring a 2-arm coincidence (ChN in anticoincidence) with 
a coplanarity restriction (HH) in the first-level trigger. 
The second-level trigger selects events with at least one track 
in each arm by exploiting DC-wire information (track finder).
Using track information, the online trigger selects $\pi \pi$ and 
$\pi K$ pairs with relative momenta 
$|Q_X| < 12~\rm{MeV}/c$ and $|Q_L| < 30~\rm{MeV}/c$. 
The trigger efficiency is about 98\% for pairs with 
$|Q_X| < 6~\rm{MeV}/c$, $|Q_Y| < 4~\rm{MeV}/c$ and $|Q_L| < 28~\rm{MeV}/c$.
Particle pairs $\pi^-{\mathrm p}$ ($\pi^+\bar{\mathrm p}$) from 
$\Lambda$ ($\bar{\Lambda}$) decay have been used for 
spectrometer calibration and $\mathrm{e}^+\mathrm{e}^-$~pairs for 
general detector calibration.

\section{Production of bound and free $\pi^- K^+$ and $\pi^+ K^-$ pairs}
\label{boundfree}

Prompt $\pi^{\mp}K^{\pm}$ pairs from 
proton-nucleus collisions are produced either directly or 
originate from short-lived (e.g. $\Delta$, $\rho$), 
medium-lived (e.g. $\omega$, $\phi$) or 
long-lived (e.g. $\eta'$, $\eta$) sources. 
Pion-kaon pairs produced directly, from short- or medium-lived 
sources, undergo Coulomb FSI resulting in 
unbound states (Coulomb pair in Fig.~\ref{Fig1_1}) or 
forming bound states ($A_{K \pi}$ in Fig.~\ref{Fig1_1}). 
Pairs from long-lived sources are practically not affected by 
Coulomb interaction (non-Coulomb pair in Fig.~\ref{Fig1_1}). 
The accidental pairs are generated via different 
proton-nucleus interactions.

The cross-section of $\pi K$ atom production 
is given by the expression \cite{NEME85}: 
\begin{equation}\label{eq:prod}
\frac{{\rm d}\sigma^{n}_A}{{\rm d}\vec p_A}=(2\pi)^3\frac{E_A}{M_A}
\left.\frac{{\rm d}^2\sigma^0_s}{{\rm d}\vec p_K {\rm d}\vec p_\pi}\right |_{
\frac{\vec p_K}{M_{K}} \approx \frac{\vec p_\pi}{M_{\pi}}}
\hspace{-1mm} \cdot \left|\psi_{n}(0)\right|^2 \:,
\end{equation}
where $\vec p_{A}$, $E_{A}$ and $M_{A}$ are the momentum, 
total energy and mass of the $\pi K$ atom in the 
laboratory (lab) system, respectively, and $\vec p_K$ and 
$\vec p_\pi$ the momenta of the charged kaon and pion 
with equal velocities. Therefore, these momenta obey 
in good approximation the relations 
$\vec p_{K}=\frac{M_{K}}{M_{A}} \vec p_{A}$ and 
$\vec p_{\pi}=\frac{M_{\pi}}{M_{A}} \vec p_{A}$. 
The inclusive production cross-section of 
$\pi K$ pairs from short-lived sources without 
FSI is denoted by $\sigma_s^0$, 
and $\psi_{n}(0)$ is the $S$-state Coulomb atom wave function  
at the origin with principal quantum number $n$.  
According to (\ref{eq:prod}), $\pi K$ atoms are only produced in 
$S$-states with probabilities $W_n=\frac{W_1}{n^3}$:      
$W_1=83.2\%$, $W_2=10.4\%$, $W_3=3.1\%$, 
$W_{n>3}=3.3\%$.

In complete analogy, the $\pi^{\mp}K^{\pm}$ Coulomb pair production 
is described in the point-like production approximation, depending  
on the relative momentum $q$ in the production point\footnote{
The quantity $q$ denotes the original c.m. relative momentum.}: 
\begin{equation}\label{eq:cross_sect_C}
\frac{{\rm d}^2\sigma_C}{{\rm d}\vec p_K {\rm d}\vec p_\pi} =
\frac{{\rm d}^2\sigma^0_s}{{\rm d}\vec p_K {\rm d}\vec p_\pi}
\hspace{-2mm} \cdot A_C(q) 
\quad \mbox{with} \quad
A_C(q) = \frac{4\pi \mu \alpha/q}
{1-\exp\left(-4\pi \mu \alpha/q\right) } \;.
\end{equation}
The Coulomb enhancement function $A_C(q)$ is the well-known 
Sommerfeld-Gamov-Sakharov factor \cite{SOM31,GAM28,SAKH91},  
$\mu$~=~109~$\rm{MeV/c^2}$ the reduced mass of 
the $\pi^{\mp}K^{\pm}$ system and $\alpha$ the fine structure constant. 
The relative production yield of atoms to Coulomb pairs~\cite{AFAN99}   
is calculated from the ratio (\ref{eq:prod}) to (\ref{eq:cross_sect_C}).

For $\pi$ and $K$ production from non-pointlike medium-lived sources, 
corrections at the percent level have been applied 
to the production cross-sections~\cite{LEDN08}. 
Strong final state elastic and inelastic interactions are 
negligible~\cite{LEDN08}.

\section{Data processing}
\label{sec:Data pro}

Recorded events have been reconstructed with the DIRAC $\pi\pi$ analysis 
software (ARIANE) modified for analysing $\pi K$ data.

\subsection{Tracking and setup tuning}
\label{ssec:Tracking}

Only events with one or two particle tracks in the DC detector  
of each arm are processed. Event reconstruction is 
performed according to the following steps:\\
1)~One or two hadron tracks are identified in the DC 
of each arm with hits in VH, HH and PSh slabs and 
no signal in ChN and Mu (Fig.~\ref{Fig2_1} and related text). 
The earliest track in each arm is used for further analysis, 
because these tracks induce the trigger signal starting 
the readout procedure.\\
2)~Track segments, reconstructed in DC, are extrapolated 
backward to the incident proton beam position in the target, 
using the transfer function of the DIRAC dipole 
magnet. This procedure provides 
approximate particle momenta and  
corresponding intersection points in MDC, SFD and IH.\\
3)~Hits are searched for around the expected SFD coordinates 
in the region $\pm 1$~cm, corresponding to 3--5~$\sigma$ 
defined by the position accuracy, taking into account 
particle momenta. This way, events are selected with 
low and medium background defined by 
the following criteria: 
the number of hits around the two tracks is $\le 4$ in each
SFD plane and  $\le 9$ in all three SFD planes. 
The case of only one hit in the region $\pm 1$~cm can occur 
because of detector inefficiency (two crossing particles, but  
one is not detected) or if two particles cross the same SFD column. 
The latter event type can be regained by double ionisation selection in 
the corresponding slab of the IH. For data collected 
in 2007 with the Pt target, criteria are different: 
the number of hits is two in the $Y$- and $U$-plane (SFD $X$-plane and 
IH, which may resolve crossing of 
only one SFD column by two particles, have not been used in 2007). 
The momentum of the positively or negatively charged particle 
is refined to match the $X$-coordinates of the tracks in DC 
as well as the SFD hits in the $X$- or $U$-plane, 
depending on presence of hits. In order to find 
the best two-track combination, the two tracks may not use 
a common SFD hit in case of more than one hit in the proper region. 
In the final analysis, the combination with the best $\chi^2$ 
in the other SFD planes is kept.

In order to improve the mechanical alignment and 
general description of the setup geometry, 
the $\Lambda$ and $\bar{\Lambda}$ particle decays into 
${\rm p}\pi^-$ and $\pi^+\bar{\rm p}$ 
are exploited~\cite{note09og,note1303,note1601}. 
By requiring the mass equality 
$M^{exp}_{\Lambda} = M^{exp}_{\bar{\Lambda}}$, 
the angles of the DC axes are modified. In the next step, 
the obtained angle between the DC axes is tuned to get 
the PDG (Particle Data group) reference $\Lambda$ mass:  
the survey value of this angle needs to be increased 
by a few $10^{-4}$~rad.  
For the data set 2007--2010, the weighted average of 
the experimental $\Lambda$ mass values is 
$M^{exp}_{\Lambda} = (1.115680 \pm 2.9 \cdot 10^{-6}) \rm{GeV}/c^2$, 
in agreement with the PDG value 
$M^{PDG}_{\Lambda} = (1.115683 \pm 6 \cdot 10^{-6}) \rm{GeV}/c^2$~\cite{pdg}. 
This confirms consistency of the setup alignment. 
The $\Lambda$ mass width in the simulated distribution tests 
how well the MC simulation reproduces the momentum and angle resolution of 
the setup. Data of each year has been investigated 
which simulated distribution -- with different widths -- fits 
best the experimental $\Lambda$ distribution. 
Simulated $\Lambda$ distributions providing a better $\chi^2$ fit to 
the data are created with a width increased by the following factors:
$1.027 \pm 0.003$ in 2007 (two SFD planes), while this increase in 
the subsequent years (three SFD planes) is not significant: 
$1.002 \pm 0.004$ (2008), 
$1.001 \pm 0.003$ (2009) and  
$1.003 \pm 0.003$ (2010). 
The difference between data and MC width could be the consequence of 
an imperfect description of the setup downstream part and 
can be removed by introducing a Gaussian smearing of 
the reconstructed momenta~\cite{ADEV14}. This technique is also used 
to evaluate the systematic error connected with 
reconstructed momentum smearing. 
Taking into account momentum smearing, 
the momentum resolution has been evaluated as  
$ \frac{dp}{p} = \frac{p_{gen} -p_{rec} } {p_{gen}}$ with $p_{gen}$ and 
$p_{rec}$ the generated and reconstructed momenta, respectively.     
Between 1.5 and 8~GeV/c, particle momenta are reconstructed with 
a relative precision from $2.8\cdot10^{-3}$ to 
$4.4\cdot10^{-3}$~\cite{note1303}. 
Relative momentum resolutions after the target are:  
$\sigma_{QX} \approx \sigma_{QY} \approx 0.36~\rm{MeV}/c$, 
$\sigma_{QL} \approx 0.94~{\rm MeV}/c$ for 
$p_{\pi K} = p_{\pi}+p_K = 5$~GeV/c and about 6\% worse values  
for $p_{\pi K} = 7.5$~GeV/c.

\subsection{Event selection}
\label{ssec:Ev_sel}

\begin{figure}[htbp]
        \begin{center}
                        
\includegraphics[width=1.0\columnwidth]{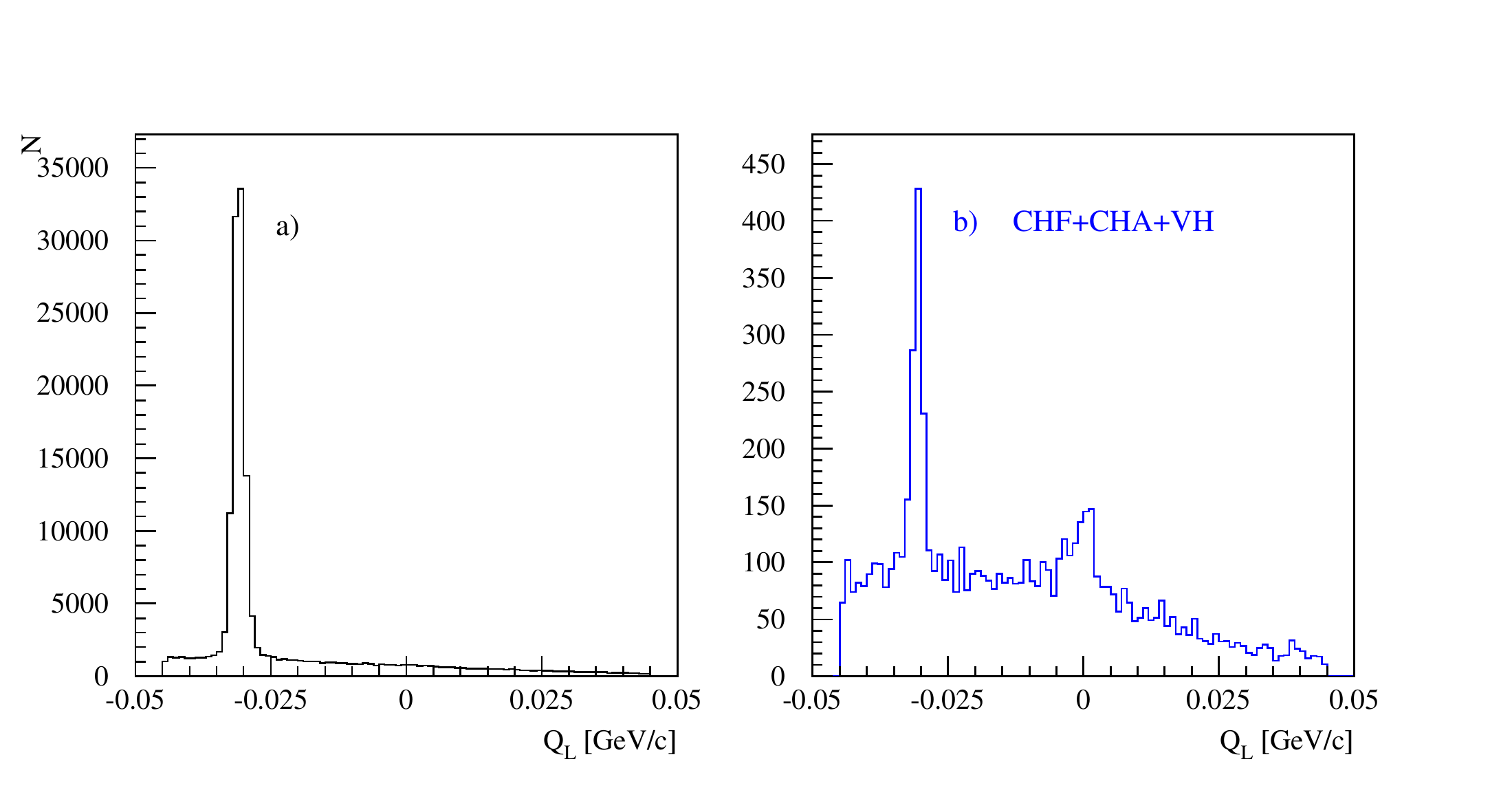}
                        \caption{
\footnotesize $Q_L$ distributions of potential $\pi^-K^+$ pairs 
before (a) and after (b) applying the selection described in the text. Events with 
positive $Q_L$ are suppressed compared to those with negative $Q_L$ 
due to lower acceptance and lower production cross-section.                     
                        }
                        \label{Fig4_1}
        \end{center}
\end{figure}

Selected events are classified into three categories: 
$\pi^-K^+$, $\pi^+K^-$ and $\pi^-\pi^+$. 
The last category is used for calibration purposes. 
Pairs of $\pi K$ are cleaned of $\pi^-\pi^+$ and $\pi^-{\rm p}$ 
background by the Cherenkov counters ChF and ChA. 
In the momentum range from 3.8 to 7~GeV/c, pions are detected 
by ChF with (95--97)\% efficiency~\cite{note1305}, 
whereas kaons and protons (antiprotons) do not produce a signal.
The admixture of $\pi^-{\rm p}$ pairs is suppressed by the 
aerogel Cherenkov detector (ChA), which records kaons  
but not protons \cite{note0907}.
By requiring a signal in ChA and 
selecting compatible time-of-flights (TOF) between the target and VH, 
$\pi^-{\rm p}$ and $\pi^-\pi^+$ pairs, contaminating $\pi^-K^+$,  
can be substantially suppressed. 
Correspondingly, the admixture of $\pi^+\pi^-$ pairs to $\pi^+K^-$ 
has also been taken into account. 
Fig.~\ref{Fig4_1} shows, after applying the selection criteria, 
the well-defined $\pi^-K^+$ Coulomb peak 
at $Q_L=0$ and the strongly suppressed peak from 
$\Lambda$ decays at $Q_L=-30$~MeV/c. 
The $Q_L$ distribution of $\pi^+K^-$ pairs is similar \cite{ADEV14}.

The final analysis sample contains only events which fulfil the following criteria:
\begin{equation}\label{eq:7c_critq}
Q_T < 4~{\rm MeV/c} \, , \, |Q_L| < 20~{\rm MeV/c} \, .
\end{equation}

\begin{figure}[htbp]
\begin{center}
\includegraphics[width=0.48\columnwidth]{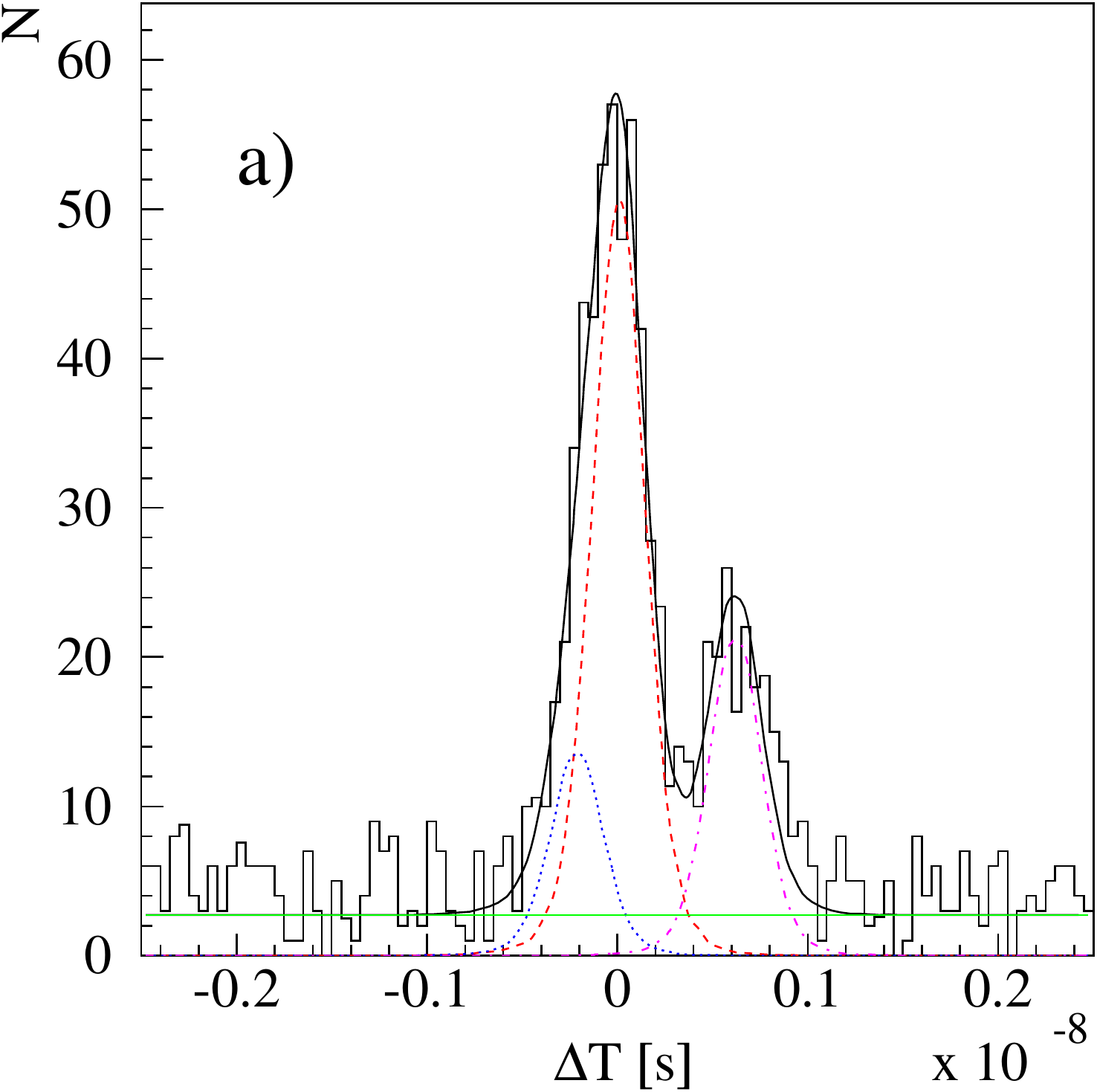} \hfill
\includegraphics[width=0.48\columnwidth]{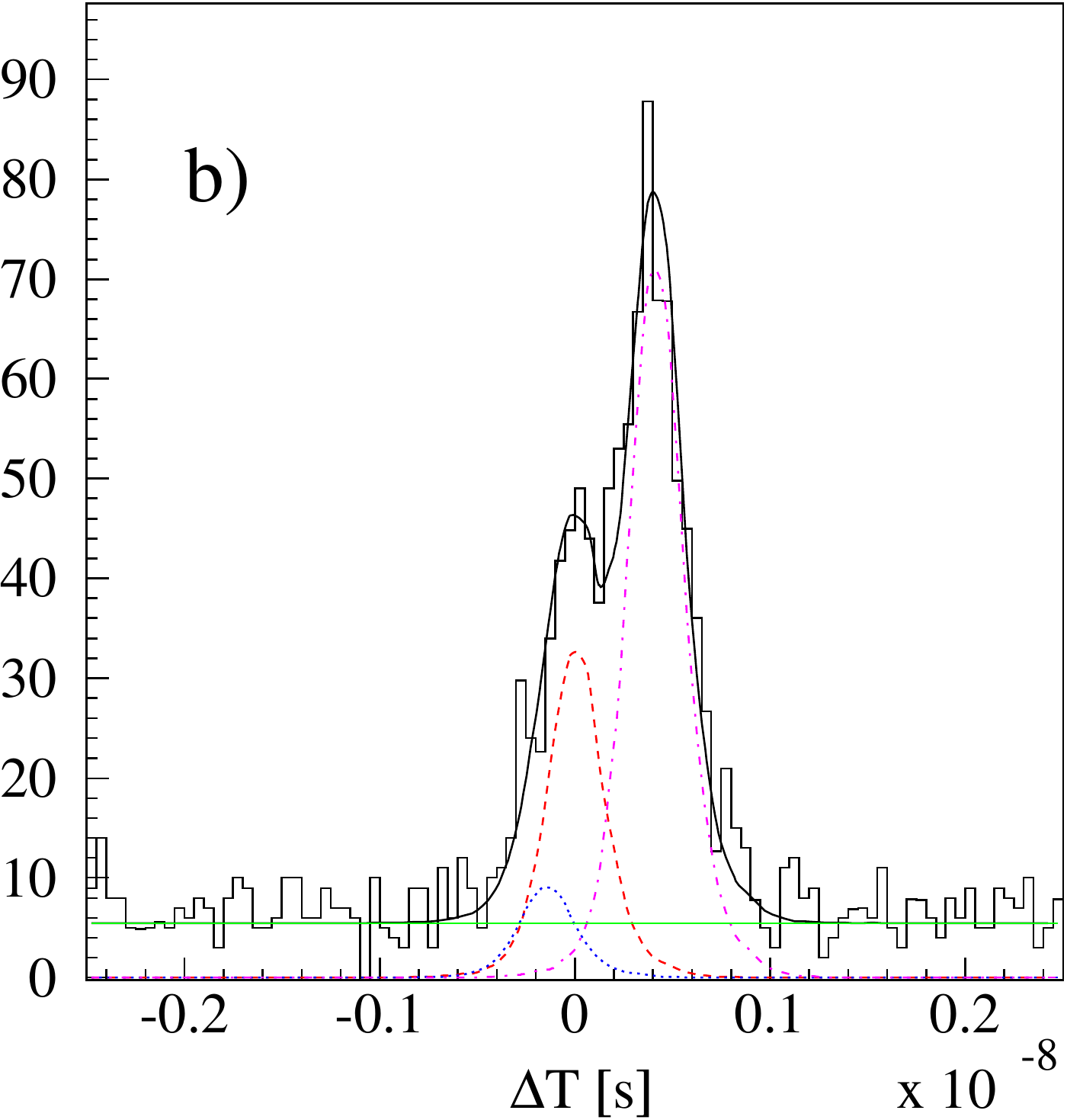}
\caption{\footnotesize Distributions over time-of-flight difference for
events with positively charged particle momenta in the intervals: 
a) (4.4--4.5) GeV/$c$; 
b) (5.4--5.5) GeV/$c$. 
Experimental data (histogram) are fitted by the sum of the distributions: 
$K^+\pi^-$ (red, dashed), 
$\pi^+\pi^-$ (blue, dotted), 
${\mathrm p}\pi^-$ (magenta, dotted-dashed) and 
accidental pairs (green, constant). 
The sum of all the fractions is shown as black solid line. }
\label{Fig4_2}
\end{center}
\end{figure}

Due to finite detector efficiency, a certain admixture of 
misidentified pairs still remains in the experimental distributions. 
Their contribution has been estimated by TOF investigation 
and accordingly been subtracted \cite{note1306}. 
Under the assumption that all positively charged particles are $K^+$, 
Fig.~\ref{Fig4_2} compares the experimental with 
the simulated TOF difference distribution for 
$\pi^-K^+$, $\pi^-\pi^+$ and $\pi^-{\rm p}$ pairs. 
Two ranges for positively charged particle momenta, 
(4.4--4.5) and (5.4--5.5) GeV/$c$, have been investigated. 

\section{Data simulation}
\label{sec:Data sim}

Since the $\pi K$ data samples consist of Coulomb, 
non-Coulomb and atomic pairs, these event types have been generated 
by MC (DIPGEN~\cite{note0711}, GEANT-DIRAC (setup simulator)).
The MC sample exceeds ten times the number of experimental events. 
The events are characterised by different $q$ distributions: 
the non-Coulomb pairs are distributed in accordance with 
phase space, while the $q$ distribution of Coulomb pairs is modified 
by the factor $A_C(q)$ (\ref{eq:cross_sect_C}). 
For atomic pairs, one needs to know the breakup position and
the lab momentum of each pair. In practice, lab momenta for 
MC events are generated in accordance with analytic formulae, 
resembling the experimental momentum distributions  
of such pairs~\cite{note0711,note1001}. 
After comparing experimental momentum spectra \cite{note1306} 
with MC distributions reconstructed by the analysis software, 
their ratio is used as event-by-event weight function for
MC events in order to provide the same lab momentum spectra  
for simulated as for experimental data. 
The breakup point, from which the ionisation occurred, 
the quantum numbers of the atomic state and the corresponding 
$q$ distribution of the atomic pair are obtained by solving 
numerically transport equations \cite{Zhab08} using total and
transition cross-sections \cite{Afan96}. The lab momenta of the atoms 
are assumed, in accordance with equation (\ref{eq:prod}), to be the same 
as for Coulomb pairs. The description of the charged particle propagation 
through the setup takes into account: 
a) multiple scattering in the target, detector planes and setup partitions, 
b) response of all detectors, 
c) additional momentum smearing and 
d) results of the SFD response analysis \cite{note1306,GORI06,BENE08} 
with influence on the $Q_T$ resolution.

\section{Data analysis}
\label{sec:Analysis}

In the analysis of $\pi K$ data, the experimental 1-dimensional distributions 
of relative momentum $Q$ and $|Q_L|$ and the 2-dimensional distributions 
($|Q_L|$, $Q_T$) have been fitted for each year and 
each $\pi K$ charge combination by simulated distributions of 
atomic, Coulomb and non-Coulomb pairs. 
Their corresponding numbers $n_A$, $N_C$ and $N_{nC}$ are free fit parameters. 
The sum of these parameters is equal to the number of analysed events. 

\begin{figure}[!htbp]
\begin{center}
\includegraphics[width=\columnwidth]{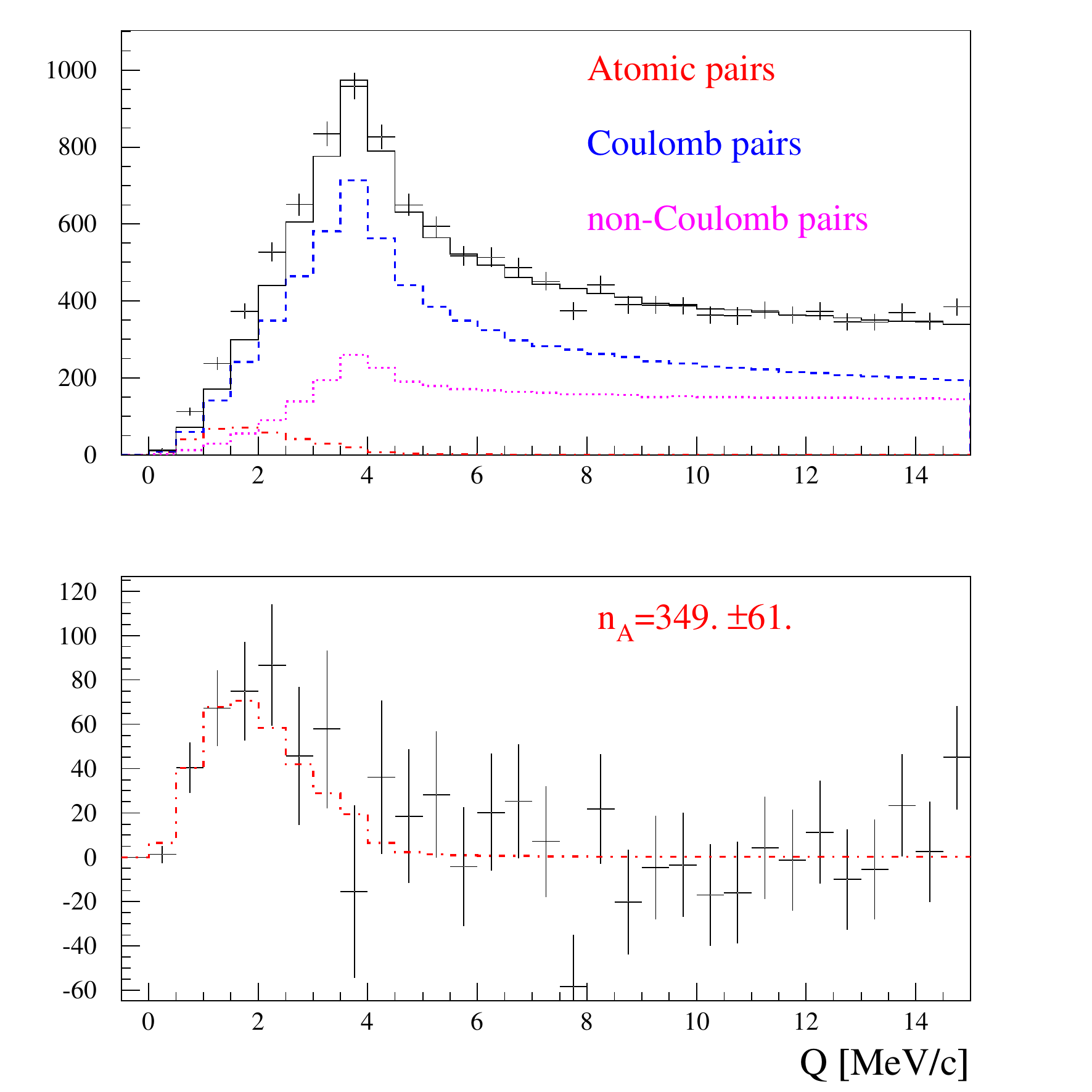}
\caption{\footnotesize Top: $Q$ distribution of experimental  
$\pi^-K^+$ and $\pi^+K^-$ pairs fitted by the sum of simulated distributions 
of atomic, Coulomb and non-Coulomb pairs.  
Atomic pairs are shown in red (dotted-dashed) and 
free pairs (Coulomb in blue (dashed) and non-Coulomb in magenta (dotted)) 
in black (solid). 
Bottom: Difference distribution between experimental and 
simulated free pair distributions compared with simulated atomic pairs. 
The number of observed atomic pairs is denoted by $n_A$. }
\label{Fig6_1}
\end{center}
\end{figure}

\begin{figure}[!htbp]
\begin{center}
\includegraphics[width=\columnwidth]{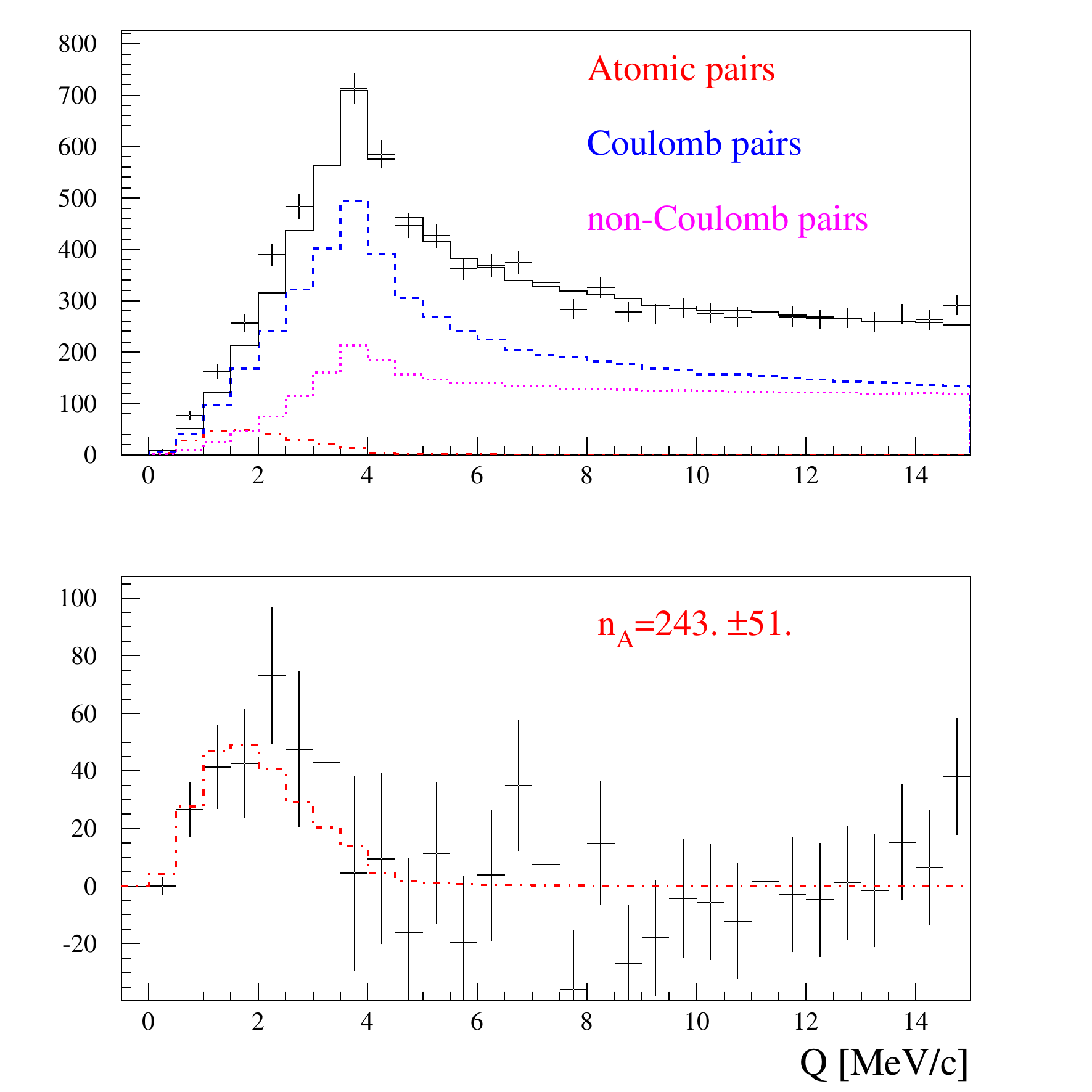}
\caption{\footnotesize Same distributions as in Fig.~\ref{Fig6_1}, 
but only for $\pi^-K^+$ pairs. }
\label{Fig6_2}
\end{center}
\end{figure}

\begin{figure}[!htbp]
\begin{center}
\includegraphics[width=\columnwidth]{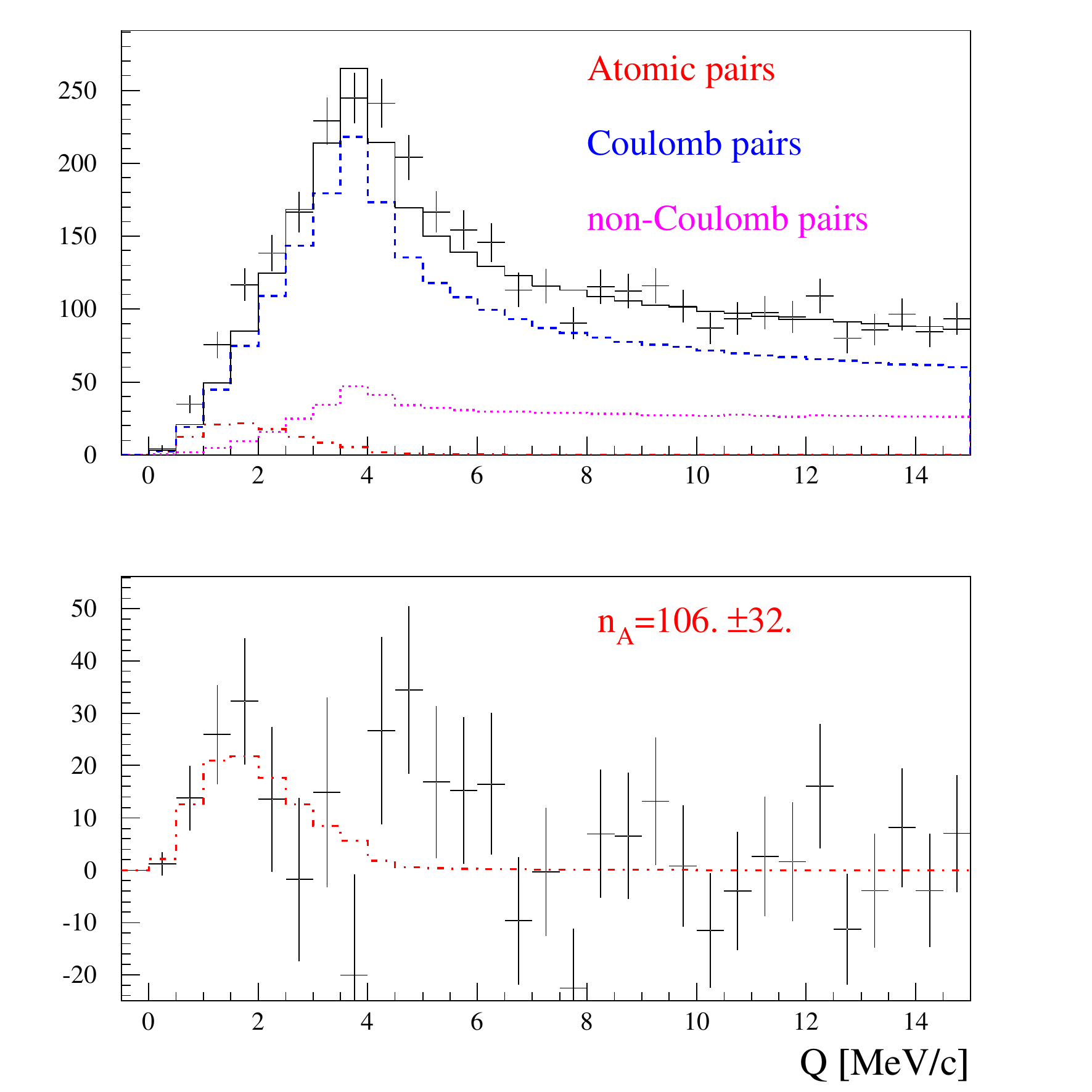}
\caption{\footnotesize Same distributions as in Fig.~\ref{Fig6_1}, 
but only for $\pi^+K^-$ pairs. }
\label{Fig6_3}
\end{center}
\end{figure}

The experimental and simulated $Q$ distributions of $\pi^- K^+$ and 
$\pi^+ K^-$ pairs are shown in Fig.~\ref{Fig6_1} (top) 
for all events with $Q_T<4$~MeV/$c$ and $|Q_L|<20$~MeV/$c$. 
One observes an excess of events above the sum of Coulomb and 
non-Coulomb pairs in the low $Q$ region, where atomic pairs
are expected. After background subtraction there is a signal at   
the level of 5.7 standard deviations, shown in Fig.~\ref{Fig6_1} (bottom): 
$n_A = 349 \pm 61$ ($\chi^2/n = 41/37$, $n=$ number of degrees of freedom), 
see Table~\ref{tab:6_1}. 
The signal shape is described by the simulated distribution of atomic pairs. 
The numbers of atomic pairs, produced in the Ni and Pt targets, are 
$n_A(\mathrm{Ni})=275\pm57$ ($\chi^2/n = 40/37$) 
and $n_A(\mathrm{Pt})=73\pm22$ ($\chi^2/n = 40/36$), respectively.
The same analysis has been performed for all $\pi^- K^+$ and $\pi^+ K^-$ pairs 
separately as presented in Fig.~\ref{Fig6_2} and Fig.~\ref{Fig6_3}. 
The $\pi^- K^+$ and $\pi^+ K^-$ atomic pair numbers are 
$n_A=243\pm51$ ($\chi^2/n = 36/37$) and 
$n_A=106\pm32$ ($\chi^2/n = 42/37$), respectively. 
The experimental ratio, $2.3\pm0.9$, between the two types of 
atom production is compatible with the ratio $2.4$ as calculated 
using FRITIOF \cite{note1505}.

In the 2-dimensional ($|Q_L|$,$Q_T$) analysis, all experimental data 
in the same $|Q_L|$ and $Q_T$ intervals have been analysed 
using simulated 2-dimensional distributions. The evaluated 
atomic pair number, $n_A=314\pm59$ ($\chi^2/n = 237/157$), corresponds to 
5.3 standard deviations and coincides with the previous analysis result. 

In Table~\ref{tab:6_1}, the results of the three analysis types 
(Ni and Pt target together) are presented for each atom type and combined.  
There is a good agreement between the results of 
the $Q$ and ($|Q_L|$,$Q_T$) analyses. 
The 1-dimensional $|Q_L|$ analysis for all $\pi K$ data yields 
$n_A=230\pm92$ ($\chi^2/n = 52/37$), which does not contradict 
the values obtained in the other two statistically more precise analyses. 

Compared to the previous investigation~\cite{ADEV09}, 
in the present work the Pt data has been analysed including 
upstream detectors. The consequence is a decrease of the statistics, 
but on the other hand an increase of the $Q_T$ resolution. 
This better resolution improves the data quality. 
Concerning the Ni target, the increase of $n_A$, compared to \cite{ADEV14}, 
is caused by optimizing the time-of-flight criteria, 
which decreases atomic pair losses for the same fraction of background 
in the final distributions.  

\begin{table}[!htb]
\small{
\caption{
Atomic pair numbers $n_A$ by analysing  
the 1-dimensional $Q$ and $|Q_L|$ distributions and 
the 2-dimensional ($|Q_L|$,$Q_T$) distribution. 
Only statistical errors are given.}
\label{tab:6_1}
\begin{center}
\begin{tabular}{|c|c|c|c|}
\hline
 Analysis & $\pi^- K^+$ & $\pi^+ K^-$ & $\pi^- K^+$ and $\pi^+ K^-$  \\
\hline
$Q$ & $243 \pm 51$ (4.7$\sigma$) & $106 \pm 32$ (3.3$\sigma$) & $349 \pm 61$ (5.7$\sigma$)\\
\hline
$|Q_L|$ & $164\pm79$ (2.1$\sigma$) & $67 \pm 47$ (1.4$\sigma$) & $230 \pm 92$ (2.5$\sigma$)\\
\hline
$|Q_L|,Q_T$ & $237\pm50$ (4.7$\sigma$) & $78 \pm 32$ (2.5$\sigma$) & $314 \pm 59$ (5.3$\sigma$)\\
\hline
\end{tabular}
\end{center}
}
\end{table}

\section{Systematic errors}
\label{sec:systematic}

The evaluation of the atomic pair number $n_A$ is affected by several 
sources of systematic errors~\cite{note1601,note1306}. Most of them are 
induced by imperfections in the simulation of the different $\pi K$ 
pairs (atomic, Coulomb, non-Coulomb) and misidentified pairs. Shape 
differences of experimental and simulated distributions in the fit 
procedure (section~\ref{sec:Analysis}) lead to biases of parameters, 
including atomic pair contribution, and finally on $n_A$. 
The influence of systematic error sources is different for the analyses  
of $Q$, ($|Q_L|,Q_T$) and $Q_L$ distributions. Table~\ref{tab:7_1} 
shows systematic errors induced by different sources.

\begin{table}[!htb]
\small{
\caption{
Systematic errors in the number $n_A$ of $\pi K$ atomic pairs.
}
\label{tab:7_1}
\begin{center}
\begin{tabular}{|p{5cm}|c|c|c|}
\hline
\rule{0pt}{12pt}Sources of systematic errors 
& $\sigma^{syst}_Q$ & $\sigma^{syst}_{Q_L}$ & $\sigma^{syst}_{|Q_L|,Q_T}$ \\[0.3em]
\hline
\rule{0pt}{12pt}Uncertainty in $\Lambda$ width correction & 0.8 & 3.0 & 2.0 \\
& & & \\
Uncertainty of multiple scattering in Ni target & 4.4 & 0.7 & 2.7 \\
& & & \\
Accuracy of SFD simulation & 0.2 & 0.0 & 0.1 \\
& & & \\
Correction of Coulomb correlation function on finite size production region & 0.0 & 0.2 & 0.1 \\
& & & \\
Uncertainty in $\pi K$ pair laboratory momentum spectrum & 3.3 & 5.4 & 7.8 \\
& & & \\
Uncertainty in laboratory momentum spectrum of background pairs & 6.6 & 1.6 & 5.4 \\
& & & \\
Total & 8.6 & 6.4 & 10.1 \\
\hline
\end{tabular}
\end{center}
}
\end{table}

\section{Conclusion}
\label{sec:conclusion}

In the dedicated experiment DIRAC at CERN, 
the dimesonic Coulomb bound states involving strangeness,  
$\pi^- K^+$ and $\pi^+ K^-$ atoms, have been observed 
for the first time with reliable statistics. 
These atoms are generated by 
a 24 GeV/$c$ proton beam, hitting Pt and Ni targets. 
In the same targets, a fraction of the produced atoms breaks up, 
leading to $\pi^- K^+$ and $\pi^+ K^-$ atomic pairs 
with small relative c.m. momenta $Q$. 
The 1-dimensional $\pi^{\mp}K^{\pm}$ analysis in $Q$ yields 
$349\pm61(stat)\pm9(syst)=349\pm62(tot)$ atomic pairs 
(5.6 standard deviations) for both charge combinations. 
Analogously, a 2-dimensional analysis in ($|Q_L|$,$Q_T$)  
has been performed with the result of 
$314\pm59(stat)\pm10(syst)=314\pm60(tot)$ atomic pairs
(5.2 standard deviations), in agreement with the former number. 

The resulting $\pi K$ atom lifetime and $\pi K$ scattering length 
from the ongoing analysis will be presented in a separate paper.

\section*{Acknowledgements}

We are grateful to R.~Steerenberg and the CERN-PS crew for 
the delivery of a high quality proton beam and 
the permanent effort to improve the beam characteristics. 
The project DIRAC has been supported by 
the CERN and JINR administration, 
Ministry of Education and Youth of the Czech Republic by project LG130131, 
the Istituto Nazionale di Fisica Nucleare and the University of Messina (Italy),  
the Grant-in-Aid for Scientific Research from 
the Japan Society for the Promotion of Science, 
the Ministry of Education and Research (Romania), 
the Ministry of Education and Science of the Russian Federation and 
Russian Foundation for Basic Research, 
the Direcci\'{o}n Xeral de Investigaci\'{o}n, Desenvolvemento e Innovaci\'{o}n, 
Xunta de Galicia (Spain) and the Swiss National Science Foundation.

\end{document}